\documentstyle[12pt,a4wide,epsf]{article}
\newlength{\extraspace}
\newlength{\extraspaces}
\newcommand{\be}{\begin{equation}
\addtolength{\abovedisplayskip}{\extraspaces}
\addtolength{\belowdisplayskip}{\extraspaces}
\addtolength{\abovedisplayshortskip}{\extraspace}
\addtolength{\belowdisplayshortskip}{\extraspace}}
\newcommand{\ee}{\end{equation}}

\newcommand{\ba}{\begin{eqnarray}
\addtolength{\abovedisplayskip}{\extraspaces}
\addtolength{\belowdisplayskip}{\extraspaces}
\addtolength{\abovedisplayshortskip}{\extraspace}
\addtolength{\belowdisplayshortskip}{\extraspace}}
\newcommand{\ea}{\end{eqnarray}}

\title{An Exact Renormalization Group analysis\\
 of $3-d$ 
 Well Developed turbulence
\footnote{To be published on Phys. Lett. {\bf B}}  }
\author{ Paolo Tomassini\\
Dipartimento di Fisica, Universit\`a di Genova \\
Istituto Nazionale di Fisica Nucleare, Sez. di Genova\\
Via Dodecaneso, 33 -I16146 Genova Italy}

\begin{document}
\maketitle

\begin{abstract}

We take advantage of peculiar properties of three dimensional
incompressible turbulence to introduce a nonstandard Exact
Renormalization Group method. 
A Galilean invariance preserving regularizing procedure is
 utilized and a
 field truncation is adopted to test the method. 
Results are encouraging:
 the energy spectrum $E(k)$  in the inertial range scales with
exponent $-1.666\pm 0.001$
 and the Kolmogorov constant $C_K$, computed for several
 (realistic) shapes of the stirring force correlator,  agrees
with
experimental data.

\end{abstract}
\par

\section{Introduction}
Exact Renormalization Group (ERG) \cite{Wetterich}\cite{M1}\cite{KU} represents a powerful tool for a
nonperturbative analysis of quantum/statistical systems. It allows us
to obtain, via an evolution equation, an effective action
$\Gamma_{\Lambda}$ in which only statistical fluctuations of scales
smaller than $1/\Lambda$ are computed. Solving this evolution equation
for  $\Lambda\rightarrow 0$ leads us to the effective action $\Gamma$,
the Legendre transform of the generating functional of connected
diagrams \cite{Wetterich}. This quite new tool has been applied to QFT 
\cite{M1}\cite{Bonini}
and critical phenomena, mainly in conjunction with a Derivative Expansion
\cite{M1}, that is a truncation of $\Gamma_{\Lambda}$ up to a
maximum number of derivatives. \par
In this letter we present  a  modified ERG method for $3-d$
turbulence and 
we explore the consequences of a simple field truncation in
solving the set of coupled differential nonlinear equations of the ERG
flow. The truncation takes into account the effects of the turbolent
diffusivity corrections as well as noise corrections due to the energy
cascade.

\section{The model}
Let us consider  an isotropic homogeneous model of incompressible  turbulence
for a Newtonian fluid of viscosity $\nu$ and density $\rho$, where the
effects of boundary/initial conditions are taken into account
by
the introduction of a stochastic noise term $f_{\alpha}(\vec{x},t)$. The ``stirred''
Navier-Stokes equations for the velocity $v_{\alpha}(\vec{x},t)$ and pressure
$p(\vec{x},t)$ are
\ba
\label{NS0}
&{ \cal D}_t v_{\alpha}(\vec{x},t)-\nu \nabla^2v_{\alpha}(\vec{x},t)+{1\over
\rho}\partial_{\alpha}
p(\vec{x},t)=f_{\alpha}(\vec{x},t)\, ,\quad  \partial_{\alpha}v_{\alpha}(\vec{x},t)=0
 \ea
\noindent where the convective derivative ${\cal 
D}_t=\left(\partial_t+ v_{\beta}\partial_{\beta}\right)$ introduces
the nonlinear term which generates the turbolent behavior. These equations
can be simplified by the introduction of a transverse
projector
$P_
{{\alpha\beta}}(\nabla)=\left(\delta_{\alpha\beta}-{\partial_{\alpha}\partial_{\beta}\over
\nabla^2}\right)$ and by the transverse convective derivative
$ D_t=\partial_t+ P(\nabla)\left(v_{\beta}\partial_{\beta}\right)$ :
\ba
\label{NS1}
D_t v_{\alpha}(\vec{x},t)-\nu \nabla^2
v_{\alpha}(\vec{x},t)=f_{\alpha}(\vec{x},t)\, .
 \ea
The stirring force $f$ is chosen be a Gaussian stochastic white noise
field which is completely determined by its two order correlator.
 Since the stirring force should mimic the instabilities
occurring near the boundaries, its correlation length must be of the
same order of the size of the system. If $L$ is such a typical length,
the correlation function 
\be
\label{corr}
<f_{\alpha}(\vec{q},\omega)f_{\beta}(\vec{q}',\omega')>=P_{\alpha\beta}(\vec{q}) F(q) (2\pi)^4
\delta^{(3)}(\vec{q}+\vec{q}')\delta(\omega+\omega')
\ee
\noindent 
\noindent must vanish for $q>>1/L$ and for $q<<1/L$. The precise shape of the spectrum $F(q)$ is not much important, nevertheless a normalization
condition due to the energy conservation  must be satisfied \cite{Gawedzky} 
$<{\cal E}>=\int {d^3q\over (2\pi)^3} F(q),$
 where  $\cal
E$ is the rate of energy dissipated by  a unit  mass of fluid.\par
To handle the statistical properties in a easy way and to throw a
connection with the usual QFT techniques, we need an action $S$ for such
model. Such action is given by the Martin, Siggia
and Rose functional  \cite{MSR}, written in terms of the transverse field $v$ and its
canonical conjugated $\hat{u}$:
\ba
\label{MSR}
&S[\hat{u},v]=\int d^3x dt\, \hat{u}_{\alpha}(\vec{x},t)D_tv_{\alpha}(\vec{x},t)-\hat{u}_{\alpha}(\vec{x},t)\nu
\nabla^2v_{\alpha}(\vec{x},t)\nonumber\\
&+{i\over 2}\int d^3x d^3y dt\,
\hat{u}_{\alpha}(\vec{x},t)P_{\alpha\beta}(\nabla)F(x-y)\hat{u}_{\beta}(\vec{y},t)\,
.
\ea
\noindent The generating
functional of connected Green function can be written as
\be
\label{gener}
{\cal W}[\hat{J},J]=-i\,  \log \left(
\int {\cal D}\hat{u}{\cal D}v\, \exp\left(iS[\hat{u},v]-i\hat{J}v-iJ\hat{u}\right)
\right)\, .
\ee
\noindent where $\hat{J}_{\alpha}(\vec{x},t)$ and
$J_{\alpha}(\vec{x},t)$ are  two auxiliary sources.  We will compute $\cal
W$ as the Legendre transform of the effective action $\Gamma[\hat{u},v]$, that is
\be
\label{LT}
{\cal W}[\hat{J},J]=\Gamma[\hat{u},v]-\hat{J}v-J\hat{u}
\ee
\noindent where $ \hat{u}$ and $v$ are determined by 
$\hat{u}_{\alpha}(\vec{x},t)=-{\delta {\cal W}\over \delta
J_{\alpha}(\vec{x},t)}$ and $ 
v_{\alpha}(\vec{x},t)=-{\delta {\cal W}\over \delta
\hat{J}_{\alpha}(\vec{x},t)} .$
The effective action  $\Gamma[\hat{u},v]$ will be obtained by means of
the ERG evolution equation.

\section{ERG in turbulence}
In this section we will recall the  main features on ERG method and we
will apply
it to the simple model of the previous section. The ERG prescription
is based on the following points.
\begin{itemize}
\item{} Consider a  quantum/statistical theory described in terms of some field $\phi$ and
action
$S[\phi]$. For notational simplicity let $\phi$ be a scalar
field. If we modify the system in such a
way that no quantum/statistical fluctuations are allowed at distances
greater than a certain scale, say, $1/\Lambda_0$ with $\Lambda_0$ very high, than the total contribution of the allowed fluctuations is very
small (and goes to zero into the limit $\Lambda_0\rightarrow
\infty$). This means that, for this modified dynamics which is  described by
some new action $S_{\Lambda_0}\equiv S+  \Delta S_{\Lambda_0}$, the effective action
$\tilde{\Gamma}_{\Lambda_0}$ computed with  $S_{\Lambda_0}$ slightly
differs from 
$S_{\Lambda_0}$ itself. 
\item{} Given $\Delta S_{\Lambda}$ one can find the evolution
equation for the effective action $\tilde{\Gamma}_{\Lambda}$  with
respect to the scale parameter $\Lambda$. Let
\be
\label{gener2}
{\cal W}_{\Lambda}
[J]=-i\,  \log \left(
\int {\cal D}\phi\, \exp\left(iS_{\Lambda}
[\phi]-iJ\phi\right)
\right)
\ee
\noindent be the generating functional of the connected Green function
of this modified theory.
Following Wetterich \cite{Wetterich}
we have:
\be
\label{evol}
\Lambda\partial_{\Lambda}\tilde{\Gamma}_{\Lambda}\vert_{\Phi}=
\Lambda\partial_{\Lambda}{\cal W}_{\Lambda}\vert_{J}=
\Lambda\partial_{\Lambda}<\Delta S_{\Lambda}[\phi]>
\ee
\noindent where $\Phi(\vec{x},t)=-{\delta {\cal W}_{\Lambda}\over \delta
J_{\alpha}(\vec{x},t)}$ is the mean field of the modified system. 
We can introduce also 
 $\Gamma_{\Lambda}[\Phi]$,  the effective
action which collapses to $S$ if fluctuations were removed:
 $\Gamma_{\Lambda}[\Phi]=
\tilde{\Gamma}_{\Lambda}[\Phi]-\Delta S_{\Lambda}[\Phi]\, .$
If $\Delta S_{\Lambda}[\phi]$ is  a quadratic term
$\Delta S_{\Lambda}[\phi]={i\over 2}\int d^dxd^dy
\phi(x)R_{\Lambda}(x-y)\phi(y),$ equation (\ref{evol}) becomes \cite{Wetterich}
\be
\label{evol3}
\Lambda\partial_{\Lambda}\Gamma_{\Lambda}\vert_{\Phi}
={i\over 2}\, tr \left(
\Lambda\partial_{\Lambda}R_{\Lambda}\cdot \left(
{1\over
{\delta^2\Gamma_{\Lambda}\over \delta\Phi\delta\Phi}+R_{\Lambda}
}\right)
\right)\, .
\ee
\noindent This is the evolution equation we are faced to solve.

\item{} Let us now consider the initial and boundary conditions for $\Gamma_{\Lambda}$. If we start
with a sufficiently high cutoff $\Lambda_0$, fluctuations are strongly
suppressed and we can safely say
$$if \quad \Lambda=\Lambda_0\, ,\, then\quad \Gamma_{\Lambda_0}[\Phi]=S[\Phi]\, .$$
Since our  goal is the computation of the ``physical'' effective action
$\Gamma[\Phi]$, that is the sum of all the 1-particle irreducible
graphs,
we require that
\be\label{lim0}
lim_{\Lambda\rightarrow 0} \, \Gamma_{\Lambda}[\Phi]=\Gamma[\Phi]\, .
\ee
\noindent Note that when $lim_{\Lambda\rightarrow
0}\Delta S_{\Lambda}=0$  equation (\ref{lim0}) is satisfied. To sum up,
if $\Delta S_{\Lambda}$ is such that \par
\indent $\quad 1)$ 
for $\Lambda\rightarrow\Lambda_0$ no quantum/statistical fluctuation
occurs
and \par
\indent 
$ \quad  2)$  $lim_{\Lambda\rightarrow
0}\Delta S_{\Lambda}=0$, \par
\noindent
 then the
solution of the evolution equation (\ref{evol3}) gives us the
effective action of the theory.

\end{itemize}
Let us now apply the method to the isotropic model of turbulence. We firstly must
look for $\Delta S_{\Lambda}$ terms to add to the MSR action in order
to satisfy the boundary conditions listed above. To do this, we need
to discuss some physical aspects underlying the turbolent system.\par
The dynamics is characterized by the presence of two typical
scales. One of them corresponds to the macroscopic size $L$ of the
system. 
$L$ is the scale at which the energy is
introduced into the system. Such energy is transferred to smaller
scales via the energy cascade
driven by the nonlinear terms \cite{Com} \cite{KO} . This energy transfer stops at very smaller scales. The second
 intrinsic scale is the typical scale $\eta_d$ at which such energy is dissipated by the
viscous term. The smaller scale $\eta_d$ is called the ``internal'' or
the
``Kolmogorov'' scale. By dimensional arguments it is easy to show (see
for example \cite{Com}) that the internal scale is related to the
physical parameters $\nu$ and $\cal E$ by
$$\eta_d\approx \left({\nu^3\over {\cal E}}\right)^{1/4}\approx {\cal
R}^{-3/4}L$$
where $\cal R$ is the Reynolds number of the model ${\cal R}=UL/\nu$,
being $U$ the typical velocity at the scale $L$. Since ${\cal R}$ is the
typical
ratio between the inertial term $(U\nabla)U$ and viscous one $\nu
\nabla^2 U$, it measures the ``strength'' of the
nonlinearity occurring at the scale $L$. \par
 Turbolent regime can be realized only if $\cal R$ is much
greater than unity. If the Reynolds number is very high, the system
develops a wide domain of the Fourier space, called ``inertial range'', in which nonlinear terms
dominate. Clearly, for
momenta
$(1/L)<<q<<(1/\eta_d)\approx {\cal R}^{3/4} (1/L)$ both boundary terms (the
stirring force in our model) and viscous terms can be neglected. In
this subrange of Fourier space we expect the system to be universal,  self
similar,
homogeneous in space and isotropic \cite{Com} .
\par
Statistical properties of the inertial range have been 
studied intensively with many methods (for a review see, for example, \cite{Com}). A  way to obtain important
information on the system is to apply a dimensional analysis
(Kolmogorov, 1941 \cite{KO}). A simple analysis shows, for example,
that the energy spectrum $
E(k)={1\over 2}{4\pi\over (2\pi)^3}tr\, <u_{\alpha}(\vec{k},t)
u_{\beta}(-\vec{k},t)>
$
 must follow a power law of the form $
E(k)=C_K{\cal E}^{2/3}k^{-{5\over 3}} $ (Kolmogorov law),
where $C_K$ is a constant (Kolmogorov constant) of the order of unity.
Clearly, being such analysis essentially a mean field approximation,
we expect the exponents of the various moments be corrected by some
anomalous
dimensions. Such corrections are linked to the intermittent behavior
of turbolent systems, that is strong nonlinear, rare,
events usually associated to instantons \cite{Com}. 
\par
Let us try to use such informations to construct a regularizing term
$\Delta S_{\Lambda}$. First of all note that if a system with a
Kolmogorov momentum scale $K_d=1/\eta_d$ is stirred at momentum scales higher
than 
$K_d$, no energy cascade develops. This can simply be inferred by looking at the Reynolds number ${\cal R}\approx
(L/\eta_d)^{4/3}<<1$. Such a system is then quite non fluctuating and the
effective action doesn't differ too much from the bare action.
This gives us the right initial condition mentioned in point $1)$.
Therefore we can 
introduce a  $\Delta S_{\Lambda}$ term which modify the characteristic size $L$ of
the system into  $\eta=1/\Lambda$, smaller than the internal scale
$\eta_d$. Since the size $L$ characterizes the force-force correlation
function only, the  $\Delta S_{\Lambda}$ term  can be chosen as a functional
that replaces the stirring force centered at the scale $L$ with a
stirring force at the  scale $\eta=(1/\Lambda)<\eta_d<<L.$
Given $F(q)$, a possible choice of  $\Delta S_{\Lambda}$ is then
\ba
\label{delta2}
& \Delta S_{\Lambda}[\phi]\equiv
{i\over 2}\int d^3x d^3y dt\,
\hat{u}_{\alpha}(\vec{x},t)R_{\Lambda,\alpha\beta}\hat{u}_{\beta}(\vec{y},t)\nonumber\\
& ={i\over 2}\int d^3x d^3y dt\,
\hat{u}_{\alpha}(\vec{x},t)P_{\alpha\beta}(\nabla)
\left(F_{\Lambda}(x-y)-F(x-y)\right)\hat{u}_{\beta}(\vec{y},t)\,
,
\ea
\noindent where $F_{\Lambda}(q)\equiv
2D_0\Lambda^{-3}h\left(q/\Lambda\right)$ has been obtained from $F(q)$ by a rescaling
$L\rightarrow\eta=1/\Lambda$. 
The parameter $D_0$ is linked, via the energy
conservation, to the rate of energy dissipation $\cal E$.  By taking into
account equation the normalization equation we have, in fact
\be
\label{d0}
 \int dq q^2 h(q/\Lambda) {D_0\over\Lambda^3}={1\over\pi^2} {\cal E}\, .
\ee
\noindent The regularized MSR action of this system $S_{\Lambda}=S+\Delta
S_{\Lambda}$
is now
\ba
\label{MSR2}
&S_{\Lambda}
[\hat{u},v]=\int d^3x dt\, \hat{u}_{\alpha}(\vec{x},t)D_tv_{\alpha}(\vec{x},t)-\hat{u}_{\alpha}(\vec{x},t)\nu
\nabla^2v_{\alpha}(\vec{x},t)\nonumber\\
&+{i\over 2}\int d^3x d^3y dt\,
\hat{u}_{\alpha}(\vec{x},t)P_{\alpha\beta}(\nabla)F_{\Lambda}
(x-y)\hat{u}_{\beta}(\vec{y},t)\, ,
\ea
\noindent and the evolution equation (\ref{evol3}) reads 
\be
\label{evol4}
\Lambda\partial_{\Lambda}\Gamma_{\Lambda}
={i\over 2} tr \left(
\Lambda\partial_{\Lambda}R_{\Lambda}\cdot \left({1\over \tilde{\Gamma}^{(2)}}\right)\vert_{\hat{u},\hat{u}}
\right)\, .
\ee
\noindent In our notation $ \left({1\over \tilde{\Gamma}^{(2)}}\right)\vert_{\hat{u},\hat{u}}
$
means the full, fields dependent, regularized propagator of the
$\hat{u}$ field.\par
Let us now check the boundary condition mentioned in point $\,2)$.
 The regularizing term (\ref{delta2}) satisfies 
 $lim_{\Lambda\rightarrow 0}\Delta
S_{\Lambda}=0 $
in the following sense.
Firstly we  take the limit
$\Lambda\rightarrow 1/L$ so that $\Delta
S_{1/L}=0$ identically. Secondly the infinite volume limit 
 $L\rightarrow \infty$ is considered. \par
Since our choice of $ \Delta S_{\Lambda}$ satisfies both initial and
boundary conditions, the effective action computed by the ERG flow
will be, in the limit $\Lambda\rightarrow 1/L\rightarrow 0$, the ``physical'' 
effective action of the model.
\par
It should be noted that our choice of the regularizing term  $
\Delta S_{\Lambda}$ is somewhat unusual. In all the previous applications
of ERG the  $ \Delta S_{\Lambda}$ term is introduced in such a way that
{\it all} the propagators of the theory are damped for $q<\Lambda$.
This is not the case in  our approach, since the correlator of the
regularized system
$<\hat{u}v>_{\Lambda}$ has support on the entire Fourier space. Note, however,
that all the
correlation functions of the {\it  physical fields}, {\it i.e.} $<vv>_{\Lambda}$
are suppressed   for $q<\Lambda$ both at the bare and at the
``dressed'' level. 
Indeed,  the correlator $<vv>_{\Lambda}\approx{\tilde{\Gamma}_{\hat{u}\hat{u}}\over |\tilde{\Gamma}_{\hat{u}v}|^2}
$ is proportional to the noise
correlation function
 $\tilde{\Gamma}_{\hat{u}\hat{u}}(p)\approx\left(F_{\Lambda}(p)+G_{\Lambda}(p)\right)$, which contains the correction of the noise term $G_{\Lambda}$.
The key point here is that the  noise $G_{\Lambda}(p)$ correction vanishes
for $p<<L$, so that the ``dressed'' noise term is damped for
$q<\Lambda$.
This is a consequence of the structure of the nonlinear terms in 3
dimensions and is deeply linked to the features of the energy
cascade \cite{PR}, where the energy is mainly transported from low to high momenta and
not {\it vice-versa} ( 2-d
turbulence and situations which give rise to backscattering \cite
{Piomelli} are here
not taken into account). 

\section{The ERG flow}
The evolution equation (\ref{evol4}) brings to an infinite system of
coupled nonlinear differential equations, one equation for every
vertex $\Gamma^{(n)}_{\Lambda}\equiv { \delta^n \Gamma_{\Lambda}\over
\delta \phi^n}\vert_{\phi=0}$. Because of  the symmetries and  of the
structure of the ERG flow, however, some restrictions are found. 
Clearly the MSR action of the systems is invariant under Galilean
transformations of an infinitesimal ``boost'' $c_{\alpha}$. Since
the regularizing action $\Delta S_{\Lambda}$ is Galilean
invariant too \cite{PR}, one finds  Ward-Takahashi  Identities (WT),
 which must be satisfied by
$\Gamma_{\Lambda}$:
\be
\label{WT}
\int d^3x dt
\left\{\left(tc_{\beta}\partial_{\beta}v_{\alpha}(\vec{x},t)-c_{\alpha}\right){\delta\over
\delta v_{\alpha}(\vec{x},t)}+
tc_{\beta}\partial_{\beta}\hat{u}_{\alpha}(\vec{x},t){\delta\over
\delta \hat{u}_{\alpha}(\vec{x},t)}
\right\}\Gamma_{\Lambda}[\hat{u},v]=0 \, .
\ee
\noindent For a more detailed analysis of WT identities see \cite{PR}.
The structure of the evolution equation poses some important
restrictions on the form of  $\Gamma_{\Lambda}[\hat{u},v]$. Indeed, the presence of the
 $\Lambda\partial_{\Lambda}R_{\Lambda}$
term in the evolution equation (\ref{evol4}) implies that no vertex
 correction with $v$ legs only can be generated by the ERG flow \cite{PR}. 
\par
Even if some restrictions on the  form of
$\Gamma_{\Lambda}[\hat{u},v]$
are found, the system of ERG differential equations is still of
infinite order.
In order to handle the ERG equations, in this work we will explore the consequences of a field
truncation
in $\Gamma_{\Lambda}$.
 The simplest truncation of 
$\Gamma_{\Lambda}$ which preserves WT identities is the following
 \ba
\label{trunc1}
&\Gamma^{trunc.}_{\Lambda}
[\hat{u},v]=\int d^3xd^3y dt\,\left\{ \delta^{(3)}(x-y)\hat{u}_{\alpha}(\vec{x},t)D_tv_{\alpha}(\vec{y},t)-\hat{u}_{\alpha}(\vec{x},t)f(\Lambda,x-y)
\nabla^2v_{\alpha}(\vec{y},t)\right\}\nonumber\\
&+{i\over 2}\int d^3x d^3y dt\,
\hat{u}_{\alpha}(\vec{x},t)P_{\alpha\beta}(\nabla)\left(F_{\Lambda}(x-y)+g(\Lambda,x-y)\right)\hat{u}_{\beta}(\vec{y},t)\,
.
\ea
\noindent This ansatz for $\Gamma_{\Lambda}$ takes into account  the
noise correction $ g(\Lambda,x-y)$
(energy cascade) as well as the damping correction $f(\Lambda,x-y)$. The
evolution equation for $\Gamma^{trunc.}_{\Lambda}$ 
reduces then to a system of two coupled nonlinear differential
equations in terms of
 $f(\Lambda,x-y)$ and $g(\Lambda,x-y)$, with the initial conditions
\be
\label{ic} 
 f(\Lambda_0,x-y)=\nu\quad  g(\Lambda_0,x-y)=0
\ee
\noindent 
 Let us start with the
computation of the effective viscosity $f(\Lambda,x-y)$, which is is defined
as
\be
\label{viscos}
-f(\Lambda,p)p^2
={\delta^2
\Gamma_{\Lambda}[\hat{u},v]\over \delta\hat{u}(\vec{p},0)\delta
v(-\vec{p},0)}\vert_{\hat{u}=v=0}\, .
\ee 
\noindent The computation of the truncated effective action
(\ref{trunc1}) can be simplified by a rescaling of the momenta
appearing into the trace, that is the loop momentum $q$, and the
external  momentum $p$. We rescale $f(\Lambda,q)$ too in the following way:
\ba
\label{v}
& x={q\over \Lambda};\quad y={p\over \Lambda};\quad
 f(\Lambda,p)=D_0^{1/3}\Lambda^{-4/3}\phi_{\Lambda}(y)
\ea
\noindent By taking into account of the explicit and implicit dependence
on the scale parameter $\Lambda$ of both the $ f(\Lambda,p)$ and
$F_{\Lambda}(p)$ functions,
the evolution equation for
$\phi_{\Lambda}(y)$ reads
\ba
\label{viscosit}
&\left(\Lambda\partial_{\Lambda}-y\partial_y-{4\over
3}\right)\phi_{\Lambda}(y)
={1\over 4\pi^2}\int
dx\int_{-1}^{1} dt\, (1-t^2)(tx^3/y-2txy-y^2)\left(3+x\partial_x\right)h(x)\times\nonumber\\
&
\left(
(x^2+y^2+2txy)\phi_{\Lambda}(x)\left(\phi_{\Lambda}\left(\sqrt{x^2+y^2+2txy}\right)(x^2+y^2+2txy)+
\phi_{\Lambda}(x)x^2
\right)
\right)^{-1}
\ea
\noindent 
The computation of the noise correction is similar. The $g(\Lambda,q)$
function is defined by
\be
\label{rumor}
g(\Lambda,p)
=-i{\delta^2
\Gamma_{\Lambda}[\hat{u},v]\over \delta\hat{u}(\vec{p},0)\delta
\hat{u}
(-\vec{p},0)}\vert_{\hat{u}=0}\, .
\ee 
\noindent As before we rescale  $g(\Lambda,q)$ in the following way
\be
\label{r}
g(\Lambda,p)=2D_0\Lambda^{-3}\chi_{\Lambda}(y)
\ee
\noindent
so that the evolution equation for $ \chi_{\Lambda}(y)$ reads
\ba
\label{rumore}
&\left(\Lambda\partial_{\Lambda}-
y\partial_y-3\right)\chi_{\Lambda}(y)=
{1\over 4\pi^2}\int dx\int_{-1}^{1} dt\, 
\left((1-t^2)y^2(x^2+y^2+2t^2x^2+3txy)
\left(3+x\partial_x\right)h(x)
\right.\nonumber\\
&\left.\left(h_{\Lambda}\left(\sqrt{x^2+y^2+2txy}\right)+\chi_{\Lambda}\left(\sqrt{x^2+y^2+2txy}\right)
\right)\right)\times\nonumber\\
&\left(
(x^2+y^2+2txy)^2\phi_{\Lambda}(x)\phi_{\Lambda}\left(\sqrt{x^2+y^2+2txy}\right)
\right. \nonumber\\
&\left.\left(\phi_{\Lambda}\left(\sqrt{x^2+y^2+2txy}\right)(x^2+y^2+2txy)+
\phi_{\Lambda}(x)x^2
\right)
\right)^{-1}
\ea
As we have seen in section 3, the Kolmogorov
scale $K_d$ is related to the viscosity and to the rate of energy
dissipation by the relation $K_d\approx({\cal E}/\nu^3)^{1/4}$. The
viscosity
$\nu$ can be expressed, then, in terms of $\cal E$ (and so $D_0$) and 
$K_d$ as $\nu\approx {\cal E}^{1/3}K_d^{-4/3}$. Remembering that the
initial cutoff $\Lambda_0$ must be chosen well beyond the dissipation
scale $K_d$, we find the initial condition
for the $\phi_{\Lambda}$ and $\chi_{\Lambda}$  functions
\be
\label{in}
\phi_{\Lambda_0}(y)\approx \left({\Lambda_0\over
K_d}\right)^{4/3}\rightarrow \infty\, ;\quad
\chi_{\Lambda_0}(y)\approx 0 \, .
\ee
\noindent 
The system of differential equations has been solved numerically with
initial conditions (\ref{in}). Universality in the inertial range means that the exact
shape
of the noise correlator should not be much important for the inertial range
statistics. We have verified the validity of  this claim, within our
approximations, by solving the ERG flow with several different noise
terms. Each noise is, of course, normalized following (\ref{d0}). In
addition, we asked the noise to vanish for null momenta,
in agreement with the $\Lambda\rightarrow 1/L\rightarrow 0$
prescription, and to 
decrease exponentially to zero for
$q>>1/L$.

\section{Results}
We report here the results of our  numerical solution for  the ERG flow. We
started at $\Lambda=\Lambda_0=1$ with initial conditions (\ref{in})
 and solved the differential equations
with an explicit finite-difference scheme. The results are obtained after a
large  number of iterations  $(\Lambda<<\Lambda_0)$. The numerical
computations show that the solutions 
approach a fixed point
$\phi_{\Lambda}(y)\rightarrow\phi(y);\chi_{\Lambda}(y)\rightarrow\chi(y)$.
For each choice of the stirring term, this fixed point doesn't depend
on the exact value of the initial condition for the viscosity
term. \par
\begin{figure}
\epsfysize=7.0in
\epsfbox[0 0 526 727]{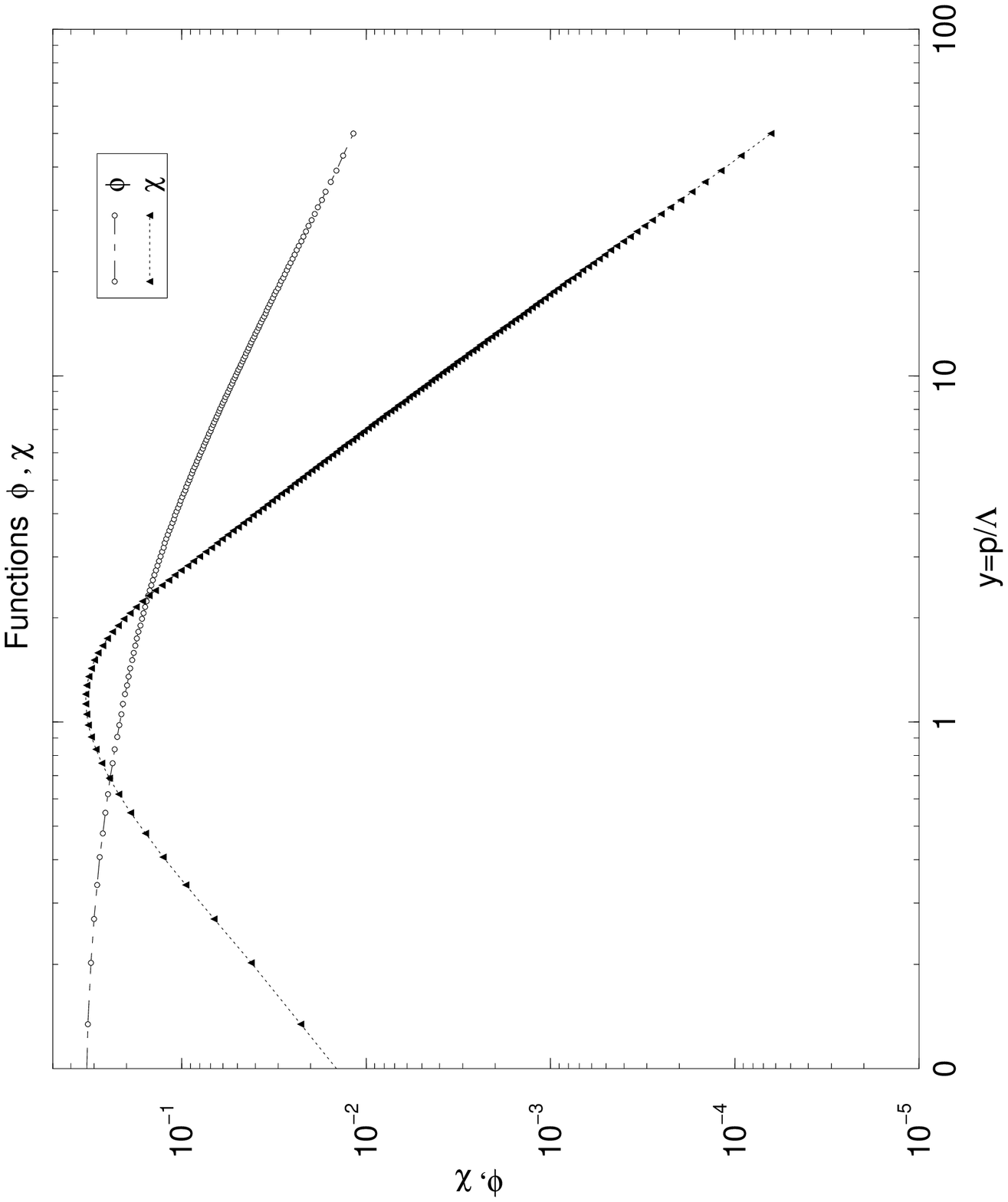}
\caption{$\phi$ and $\chi$ functions}\label{fig:2}
\end{figure}

A possible form of the stirring term is
$h(pL)=L^2p^2\,\exp\left(-L^2p^2\right)$.
In figure  (\ref{fig:2}) we report the results for the fixed point
functions $\phi(y)$ and $\chi(y)$ obtained with this noise.\par
\begin{figure}
\epsfysize=7.0in
\epsfbox[0 0 526 727]{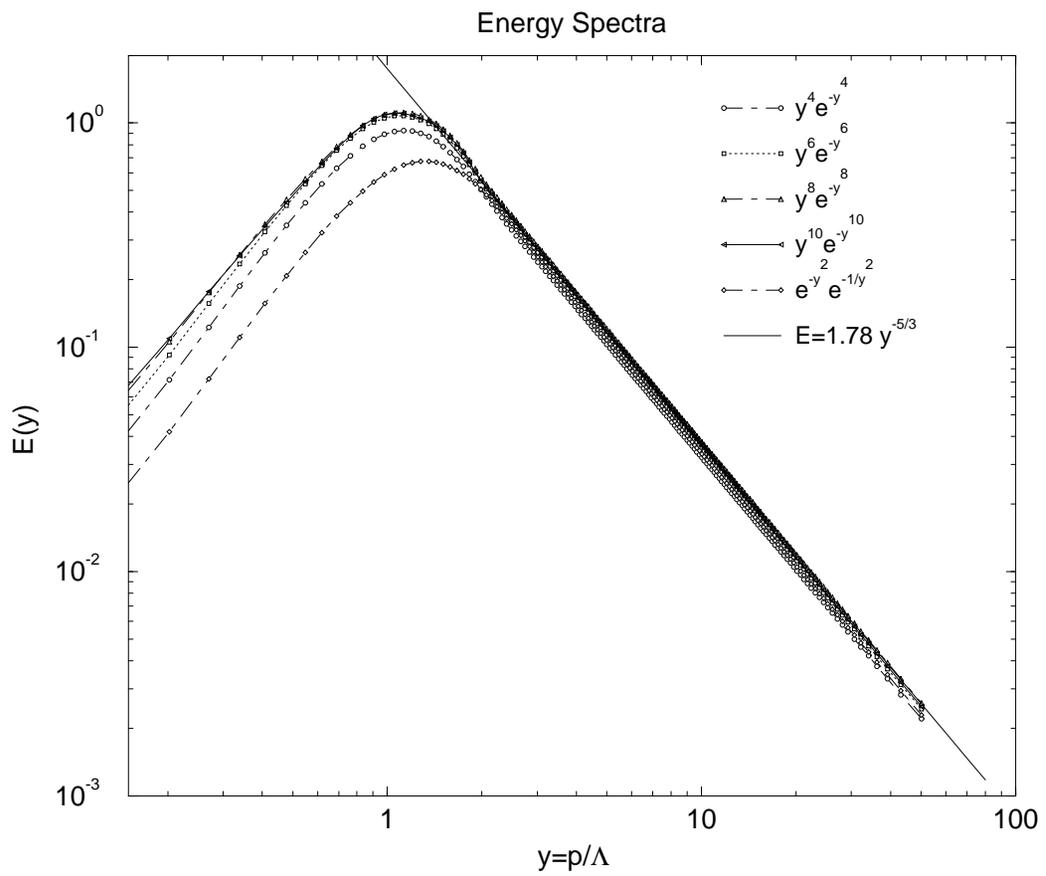}
\caption{Energy spectra for various noise terms}\label{fig:4}
\end{figure}

The functions $\phi(y)$ and $\chi(y)$ are computed  in the range
$10^{-1}<y={p\over\Lambda}<10^2$. Clearly the subrange $y>>1$
corresponds to the inertial range. As it was expected, in this region
the
adimensionalized
effective viscosity $\phi$ and induced noise $\chi$  functions reach a
 scale-invariant regime, so that we parametrize them as
\be
\label{scale}
\phi(y)_{y>>1}=\sigma y^{-4/3+\alpha}\, ; \quad
\chi(y)_{y>>1}=\gamma y^{-3+\beta}\, ,
\ee
\noindent where $\alpha$ and $\beta$ ($\alpha\approx \beta\approx +0.37$) are anomalous exponents.
 The first check for the validity of our approach is
 the computation of the energy spectrum. The
correlation function at equal long time $t$ is computed as 
\ba
&<u_{\alpha}(\vec{p},t)u_{\beta}(-\vec{p},t)>
=P_{\alpha\beta}(\vec{p})
{2D_0 \left(\Lambda^{-3}h(p/\Lambda)+g(\Lambda,p)\right)\over
f(\Lambda,p)p^2}
\ea
\noindent By taking into account  (\ref{v}) and (\ref{r}),
in the inertial range region ($y>>1$) we obtain
\be
\label{spe2}
E(p)={1\over 2\pi^2}p^2{\chi(p/\Lambda)\over
\phi(p/\Lambda)p^2}\Lambda^{-5/3}\left({D_0}\right)^{2/3}=
\left(
{1\over 2\pi^2}\left(
{\gamma\over \sigma}\right)\left({D_0\over {\cal
E}}\right)^{2/3}\right){\cal E}^{2/3} p^{-5/3}
\left({p\over \Lambda}\right)^{\beta-\alpha}
\ee
\noindent 
so that $
C_K=\left(
{1\over 2\pi^2}\left({\gamma\over \sigma}\right)\left({D_0\over {\cal
E}}\right)^{2/3}\right) . $ 
From data of figure
(\ref{fig:2}) we can obtain the plot of the Energy spectrum by a simple
ratio
between the $\chi $ and $\phi$ data normalized by $ {1\over 2\pi^2}\left({D_0\over {\cal
E}}\right)^{2/3} $ . 
In the plot (\ref{fig:4}) we have resumed the spectra obtained with
several 
different noise shapes, while in the table $T1$  we list the values of
$C_K$ and of the slope $S=-5/3+(\beta-\alpha)$ obtained with each noise.

\begin{center}
\label{t}
\begin{tabular}{|c|c|c||c|c|c|}
\hline
 Noise & $S$  ($\pm 0.001$) & $C_K$ ($\pm 0.002 $) 
& Noise & $S$ ($\pm 0.001$) &  $C_K$ ($\pm 0.002 $) \\ \hline
$ x^2\, e^{-x^2}$   & -1.666 & 1.124 &
$ x^4\, e^{-x^4}$   & -1.666 & 1.146  \\
$ x^2\, e^{-x^4}$   & -1.667 & 1.267 &
$ x^6\, e^{-x^6}$   & -1.666  & 1.660 \\
$ x^{2}\, e^{-x^{6}}$   & -1.666 & 1.417 &
$ x^8\, e^{-x^8}$   &-1.666 & 1.767 \\
$ e^{-1/x^2-x^2}$   & -1.667 & 1.489 &
$ x^{10}\, e^{-x^{10}}$& -1.666 & 1.784 \\
$ x^4e^{-12}$   & -1.667 & 1.624 &
$ x^{12}\, e^{-x^{12}}$& -1.666 & 1.785 \\
\hline
\end{tabular}
\end{center}
\centerline{Table T1: Values of the slope $S$ and of the Kolmogorov constant
$C_K$ for various stirring forces}

\section{Comments}
To sum up, in order to apply the ERG  method to turbulence, we
proposed
an unconventional, symmetry preserving, regularization procedure.
In addition, we don't need to use the  noise correlator of
the form $F(q)\approx q^{-y}$ which is used in quite all the previous
RG approaches \cite{DM}\cite{FSN}\cite{FF}\cite{YO} (an exception is
given by Brax's work \cite{Brax}). Such form is clearly unphysical, as
it is widely stressed \cite{Com}\cite{E1}. \par
To test the method a field
truncation is applied to the ERG flow.
The results obtained by this quite crude approximation are encouraging.
As it  is clear from the table and figure (\ref{fig:4}), the second
order
inertial range statistics  shows the 
required scale-invariance with  scale exponents compatible with $-5/3$
( for the second order statistics the anomalous exponent, if any, must be very small \cite{PO} ) as well as
strong independence with respect to large scale energy input details.
The values of the $C_K$ constant approach the value $C_K\approx 1.78
$ in the limit of very narrow noise form (for experimental results see
for example \cite{PO}).\par
 This approach is worth to be further
analyzed, for instance, 
with higher orders field truncations or, more likely, with nonpolinomial
approximations schemes.  \par
In a next paper \cite{PR} the complete analysis of such procedure will
be  reported and different approximation schemes will be considered.

\section*{Acknowledgments}
\noindent We would thank G. Cella, R. Collina, G. Curci, 
E. Guadagnini and F. Toschi
for useful discussions and technical supports.

\newpage
\thispagestyle{empty}

\end{document}